      \documentstyle[12pt,graphicx]{article}                
\begin{document}
\baselineskip 18pt
\def\today{\ifcase\month\or
 January\or February\or March\or April\or May\or June\or
 July\or August\or September\or October\or November\or December\fi
 \space\number\day, \number\year}
\def\thebibliography#1{\section*{References\markboth
 {References}{References}}\list
 {[\arabic{enumi}]}{\settowidth\labelwidth{[#1]}
 \leftmargin\labelwidth
 \advance\leftmargin\labelsep
 \usecounter{enumi}}
 \def\newblock{\hskip .11em plus .33em minus .07em}
 \sloppy
 \sfcode`\.=1000\relax}
\let\endthebibliography=\endlist
\def\lsim{\ ^<\llap{$_\sim$}\ }
\def\gsim{\ ^>\llap{$_\sim$}\ }
\def\r2{\sqrt 2}
\def\beq{\begin{equation}}
\def\eeq{\end{equation}}
\def\beqn{\begin{eqnarray}}
\def\eeqn{\end{eqnarray}}
\def\rmuu{\gamma^{\mu}}
\def\rmud{\gamma_{\mu}}
\def\PL{{1-\gamma_5\over 2}}
\def\PR{{1+\gamma_5\over 2}}
\def\sinW2{\sin^2\theta_W}
\def\AEM{\alpha_{EM}}
\def\mul{M_{\tilde{u} L}^2}
\def\mur{M_{\tilde{u} R}^2}
\def\mdl{M_{\tilde{d} L}^2}
\def\mdr{M_{\tilde{d} R}^2}
\def\mz2{M_{z}^2}
\def\c2b{\cos 2\beta}
\def\au{A_u}         
\def\ad{A_d}
\def\cob{\cot \beta}
\def\v#1{v_#1}
\def\tb{\tan\beta}
\def\epem{$e^+e^-$}
\def\KK{$K^0$-$\bar{K^0}$}
\def\wi{\omega_i}
\def\xj{\chi_j}
\def\Wmu{W_\mu}
\def\Wnu{W_\nu}
\def\m#1{{\tilde m}_#1}
\def\mH{m_H}
\def\mw#1{{\tilde m}_{\omega #1}}
\def\mx#1{{\tilde m}_{\chi^{0}_#1}}
\def\mc#1{{\tilde m}_{\chi^{+}_#1}}
\def\mwi{{\tilde m}_{\omega i}}
\def\mxi{{\tilde m}_{\chi^{0}_i}}
\def\mci{{\tilde m}_{\chi^{+}_i}}
\def\mz{M_z}
\def\sw{\sin\theta_W}
\def\cw{\cos\theta_W}
\def\cb{\cos\beta}
\def\sb{\sin\beta}
\def\rwi{r_{\omega i}}
\def\rxj{r_{\chi j}}
\def\rfp{r_f'}
\def\Kik{K_{ik}}
\def\Fq2{F_{2}(q^2)}
\def\f{\({\cal F}\)}
\def\d1{{\f(\tilde c;\tilde s;\tilde W)+ \f(\tilde c;\tilde \mu;\tilde W)}}
\def\tw{\tan\theta_W}
\def\sec2w{sec^2\theta_W}

\begin{titlepage}

\begin{center}
{\large {\bf Corrections to the Higgs Boson Masses and Mixings from  
Chargino, W and Charged Higgs Exchange Loops and Large CP Phases}}\\
\vskip 0.5 true cm
\vspace{2cm}
\renewcommand{\thefootnote}
{\fnsymbol{footnote}}
 Tarek Ibrahim$^{a,b}$ and Pran Nath$^{b,c}$  
\vskip 0.5 true cm
\end{center}

\noindent
{a. Department of  Physics, Faculty of Science,
University of Alexandria,}\\
{ Alexandria, Egypt\footnote{: Permanent address of T.I.}}\\ 
{b. Department of Physics, Northeastern University,
Boston, MA 02115-5000, USA\footnote{: Permanent address of P.N.}} \\
{c. Physikalisches Institut, Universit\"at Bonn, 
Nussallee 12, D-53115 Bonn, Germany}\\
\vskip 1.0 true cm
\centerline{\bf Abstract}
\medskip
One loop contributions to the Higgs boson masses and mixings from the
chargino sector consisting of the chargino, the W, and the charged 
Higgs boson ($\chi^+-W-H^+$) exchanges and including the effects of large 
CP violating phases are computed. It is found that the chargino sector
makes a large contribution to the mixings of the CP even and the CP 
odd Higgs sectors through the induced one loop effects and may even 
dominate the mixing generated by the stop and the sbottom sectors. 
Effects of the chargino sector contribution to the Higgs boson masses are 
also computed. It is found that the sum of the $\chi^+-W-H^+$ exchange 
contribution lowers the lightest Higgs boson mass and worsens the fine tuning
problem implied by the LEP data. Phenomenological implications of these
results are discussed.
\end{titlepage}

\section{Introduction}
In the Supersymmetric Standard Model [MSSM]\cite{applied} the 
loop corrections to the effective potential\cite{coleman,arnowitt1}
make very important contributions to the
Higgs masses\cite{ellis0}. Thus in the absence of the loop corrections 
the lightest Higgs mass satisfies the inequality $m_h<M_Z$ 
already in contradiction with the current lower limits from LEP.
However, with the inclusion of radiative corrections 
the lightest Higgs mass can be lifted above 
$M_Z$. The major correction to the lightest Higgs mass comes from 
the stop exchange contribution and analyses have been
extended to include sbottom exchange, the leading two loop contributions
  and several other refinements\cite{ellis0}. 
These refined analyses are then expected to yield Higgs masses
accurate to a level of 1-2 GeV. In this paper we give an 
analysis of the effect of CP violating phases 
on the Higgs masses including the effects from the exchange of
charginos, the W boson, and the charged Higgs. These contributions
modify very significantly the mixings between the CP even and the
CP odd Higgs sectors from the  CP violation effects  arising from 
the stop and sbottom exchange loops. Below we first describe the
motivation that leads us to the analysis of the effects of large
CP phases in this context.

 As is well known in SUSY models with 
softly broken supersymmetry new sources of CP violation arise
from the soft SUSY breaking parameters which are in general
complex. The natural size of these phases is large, typically
O(1), and an order of magnitude estimate shows that they can
lead to the electric dipole moment (edm) of the electron and of the
neutron in excess of the current experiment which for the 
electron is  $|d_e|<4.3 \times 10^{-27} ecm$\cite{commins} 
and for the neutron is 
$|d_n|<6.3 \times 10^{-26} ecm$\cite{harris}. 
There are several solutions suggested
in the literature to overcome this problem. Thus one
possibility suggested early on was that the phases could
be small\cite{ellis} while another possibility is that the
emds are suppressed because of the heaviness of the 
sparticle spectrum that enters in the loops that contribute
to the edms\cite{na}. Each of these possibilities is not very 
attractive. Thus the assumption of small phases constitutes
a fine tuning while the assumption that the sparticle 
spectrum is heavy may put the sparticles even beyond the
reach of the Large Hadron Collider (LHC). Another possibility
suggested recently seems more encouraging, i.e., that
 the CP phases have their natural sizes
 O(1) and the compatibility with the experimental edm constraints
occurs because of internal cancellations among the various
contributions to the edms\cite{in1}.  In this scenario 
one can have a light sparticle spectrum which would be 
accessible at colliders. This suggestion has been investigated
in considerable further detail\cite{in2}. We note in passing 
that in several recent works an assumption has been made
in order to overcome the edm problem 
that the entire CP violating effect in the SUSY sector 
arises from the phase of the trilinear coupling in the
third generation sector. In the absence
of a symmetry that would guarantee the vanishing of the 
CP violating phase of $\mu$, the phases of the gauginos,
 and  the phases 
in the first two generations such an assumption is equivalent to 
fine tuning all these phases to zero. 
It is appropriate to note
that the cancellation mechanism also requires an adjustment of phases
of currently unknown origin. One hopes that  eventually when
one learns how supersymmetry breaks in string theory, that such
a breaking will determine the phases and pick out the right mechanism
of the three listed above, or even something entirely new may come up 
to solve the edm problem in SUSY theory. In the analysis of this paper
we adopt the view that the phases are indeed large and that it is the 
cancellation mechanism which provides the correct solution.

In the presence of
large CP phases the effects of such phases on low energy 
phenomenon can be very significant and analyses have been 
carried out to investigate their effects on dark matter, 
on $g_{\mu}-2$ and on  other low energy processes. 
One area of special interest to us 
here where the presence of large CP violating phases have
been investigated is the Higgs sector\cite{pilaftsis}.
It was pointed out in Ref.\cite{pilaftsis} 
that the presence of CP violating phases in the soft SUSY
breaking sector will induce CP violating effects in the
Higgs sector allowing a mixing of the CP even and CP odd
Higgs sectors. One consequence of this mixing is that the
Higgs mass matrix no longer factors  into a $2\times 2$
CP even Higgs mass matrix times a  CP odd Higgs sector.
Consequently the  diagonalization of the neutral Higgs mass matrix
involves the diagonalization of a $3\times 3$ matrix 
in MSSM reflecting the mixing between the two CP even and one 
CP odd Higgs fields. Effects of CP violating phases 
arising from the exchange of the stops and sbottoms were computed 
in Ref.\cite{pilaftsis,pilaftsis2,demir,drees, carena}. 
In this 
work we include also the corrections due to the  exchange of 
the charginos, the W boson, and the charged Higgs. 
The  chargino exchange brings in an additional
CP violating phase which is the phase of the SU(2) gaugino
mass $\tilde m_2$.
 
 To define notation we recall that in mSUGRA\cite{chams,applied}
 the low
energy parameters are given by $m_0$, $m_{\frac{1}{2}}$,
$A_0$, $\tan\beta$ and $\theta_{\mu}$ where $m_0$ is the universal
scalar mass,  $m_{\frac{1}{2}}$ is the universal  gaugino
mass, $A_0$ is the universal trilinear coupling, 
$\tan\beta =\frac{v_2}{v_1}$ is the ratio of the Higgs VEVs,
where the VEV of $H_2$  gives mass to the up quarks and 
the VEV of $H_1$ gives mass to the down quarks and the leptons, 
and $\theta_{\mu}$ is the phase of the Higgs mixing parameter $\mu$,
 where the parameter $\mu$ is determined via radiative breaking
of the electro-weak symmetry. 
In mSUGRA there are
only two independent CP phases which can be taken to be
$\theta_{\mu_0}$ the phase of $\mu_0$ (the value of $\mu$ at 
the GUT Scale) and $\alpha_{A_0}$,
the phase of $A_0$. In this paper, however, we shall carry out
the analysis for the more general case of the MSSM. In this case
we shall treat the phases $\alpha_{A_t}$, $\alpha_{A_b}$,
$\theta_{\mu}$, and the phases $\xi_i$ (i=1,2,3) of the
$SU(3)\times SU(2)\times U(1)$ gaugino masses 
$\tilde m_i$ (i=1,2,3) all taken  at
the electro-weak scale to be independent. 
In MSSM the 
 Higgs sector at the one loop level is described by the 
scalar potential\cite{applied} 
\beq
V(H_1,H_2)=V_0+\Delta V\nonumber\\
\eeq
In our analysis we use the renormalization group improved 
effective potential where 
\beqn
V_0=m_1^2 |H_1|^2+m_2^2|H_2|^2 +(m_3^2 H_1.H_2 + H.C.)\nonumber\\
+\frac{(g_2^2+g_1^2)}{8}|H_1|^4+
\frac{(g_2^2+g_1^2)}{8}|H_2|^4
-\frac{g_2^2}{2}|H_1.H_2|^2
+\frac{(g_2^2-g_1^2)}{4}|H_1|^2|H_2|^2 
\eeqn
where $m_1^2=m_{H_1}^2+|\mu|^2,~~~~m_2^2=m_{H_2}^2+|\mu|^2,
~~~~m_3^2=|\mu B|$
and $m_{H_{1,2}}$ and $B$ are the soft SUSY breaking parameters,
and $\Delta V$ is the one loop correction to the effective 
potential and is given by\cite{coleman}

\beq
\Delta V=\frac{1}{64\pi^2}
 Str(M^4(H_1,H_2)(log\frac{M^2(H_1,H_2)}{Q^2}-\frac{3}{2}))
\eeq
where $Str=\sum_i C_i (2J_i+1)(-1)^{2J_i}$
where the sum runs over all particles with spin $J_i$ 
and $C_i(2J_i+1)$ counts the degrees of freedom of the particle i,
and Q is the running scale. In the evaluation
of $\Delta V$ one should include the contributions of all of the 
fields that enter in MSSM. This includes the Standard Model fields 
and their superpartners, the sfermions, the higgsinos and the 
gauginos\cite{arnowitt1}. The one loop corrections
to the effective potential make significant contributions 
to the minimization conditions\cite{arnowitt1}.

  As observed in Ref.\cite{pilaftsis} as a consequence of the CP 
violating effects in the one loop effective
potential the Higgs VEVs develop an induced CP violating phase
through the minimization of the effective potential. One can
parametrize this effect by the CP phase $\theta_H$ where
\beqn
(H_1)= \left(\matrix{H_1^0\cr
 H_1^-}\right)
 =\frac{1}{\sqrt 2} 
\left(\matrix{v_1+\phi_1+i\psi_1\cr
             H_1^-}\right)\nonumber\\
(H_2)= \left(\matrix{H_2^+\cr
             H_2^0}\right)
=\frac{e^{i\theta_H}}{\sqrt 2} \left(\matrix{H_2^+ \cr
             v_2+\phi_2+i\psi_2}\right)
\eeqn
The non-vanishing of the phase $\theta_H$ can be seen by
looking at the minimization of the effective potential.
For the present case with the inclusion of CP violating effects 
  the variations with respect to
the fields $\phi_1, \phi_2, \psi_1,\psi_2$ give the following

\beq
-\frac{1}{v_1}(\frac{\partial \Delta V}{\partial \phi_1})_0=
m_1^2+\frac{g_2^2+g_1^2}{8}(v_1^2-v_2^2)+m_3^2 \tan\beta \cos\theta_H
\eeq

\beq
-\frac{1}{v_2}(\frac{\partial \Delta V}{\partial \phi_2})_0=
m_2^2-\frac{g_2^2+g_1^2}{8}(v_1^2-v_2^2)+m_3^2 cot\beta \cos\theta_H
\eeq

\beq
\frac{1}{v_1}(\frac{\partial \Delta V}{\partial \psi_2})_0=
m_3^2 \sin\theta_H= \frac{1}{v_2}
(\frac{\partial \Delta V}{\partial \psi_1})_0
\eeq
where the subscript 0 means that the quantities are evaluated
at the point $\phi_1=\phi_2=\psi_1=\psi_2=0$.
As noted in Ref.\cite{demir} only one of the two equations 
in Eq.(7) is independent.

\section{Chargino,W and charged Higgs contributions}
The contribution of the stop and of the sbottom exchange contributions
have been discussed at great length in the literature\cite{ellis0}. 
More recently these analyses have been extended to take account of
the CP violating effects arising from the soft SUSY breaking 
parameters in these sectors\cite{pilaftsis,pilaftsis2,demir,drees,carena}.
We have reanalysed the stop and 
sbottom contributions with CP violating effects and these 
results are listed in Appendix A where we also compare our 
results with the previous analyses. The main focus of this
work, however, is to compute the contributions of the 
chargino loops to the Higgs masses. 
The charginos, the W and the charged Higgs boson form a sub-sector
as it is the splittings among these particles that leads to a 
non-vanishing contribution to the one loop effective potential.
The one loop correction from this sector is given by 
\beqn
\Delta V(\chi^+,W, H^+)=\frac{1}{64\pi^2}
(\sum_{a=1,2}(-4)M_{\chi_a^+}^4(log\frac{M_{\chi_a^+}^2}{Q^2}-\frac{3}{2})
+6 M_W^4 (log\frac{M_W^2}{Q^2}-\frac{3}{2})\nonumber\\
+ 2 M_{H^+}^4 log(\frac{M_{H^+}^2}{Q^2}-\frac{3}{2}))
\eeqn
 The chargino mass matrix is given by 
\beq
M_C=
\left(\matrix{\tilde m_2 & g_2 H_2^0\cr
g_2 H_1^0 & \mu}\right)
\eeq
where $\mu=|\mu|e^{i\theta_{\mu}}$  
and $\tilde m_2=|\tilde m_2|e^{i\xi_2}$.
For the purposes of the analysis it is more convenient to deal
with the matrix $M_CM_C^{\dagger}$ where
\beq
M_CM_C^{\dagger}=
\left(\matrix{|\tilde m_2|^2+g_2^2 |H_2^0|^2 &
 g_2 (\tilde m_2 H_1^{0^*} +\mu^* H_2^0)\cr
g_2 (\tilde m_2^* H_1^0 +\mu H_2^{0^*}) & |\mu|^2+ g_2^2 |H_1^0|^2}\right)
\eeq
The chargino eigen values are given by

\beqn
M_{\chi_{1,2}^+}^2 =\frac{1}{2}
[|\tilde m_2|^2+|\mu|^2 + g_2^2 (|H_2^0|^2+|H_1^0|^2)]\nonumber\\
\pm \frac{1}{2}[(|\tilde m_2|^2-|\mu|^2+g_2^2(|H_2^0|^2-|H_1^0|^2))^2
+4 g_2^2|\tilde m_2H_1^{0*}+\mu^* H_2^0|^2]^{\frac{1}{2}}
\eeqn
We note that in the supersymmetric limit $M_{\chi_{1,2}^+}$=
$M_{H^+}=M_W$ and the loop correction Eq.(8) vanishes.
Further, as we will discuss later the inclusion of the W and the
$H^+$ exchange along with the chargino exchange is also needed to
achieve an approximate Q independence of the corrections to 
the Higgs masses and mixings from this sector. 
In this sense $M_{\chi_{a}^+}, H^+$ and  W form a sub-sector
and that is the reason for considering this set in Eq.(8).
With the inclusion of the stop and the sbottom 
contributions (see Appendix A)  and of the chargino 
contributions one finds that $\theta_H$ is determined by
the equation

\beqn
m_3^2 \sin\theta_H =\frac{1}{2} \beta_{h_t} |\mu| |A_t| \sin\gamma_t
f_1(m_{\tilde t_1}^2, m_{\tilde t_2}^2)
+\frac{1}{2}  \beta_{h_b}|\mu| |A_b| \sin\gamma_b
f_1(m_{\tilde b_1}^2, m_{\tilde b_2}^2)\nonumber\\
-\frac{g^2_2}{16\pi^2}|\mu| |\tilde m_2| \sin\gamma_2
f_1(m_{\tilde \chi_1}^2, m_{\tilde \chi_2}^2)
\eeqn
where 
\beq
 \beta_{h_t}=\frac{3 h_t^2}{16\pi^2}, 
~~ \beta_{h_b}=\frac{3 h_b^2}{16\pi^2};~~
\gamma_t=\alpha_{A_t} + \theta_{\mu}, ~~
\gamma_b=\alpha_{A_b} + \theta_{\mu},~~ 
\gamma_2=\xi_2 + \theta_{\mu}
\eeq
and $f_1(x,y)$ is defined by
\beq
f_1(x,y)=-2+log\frac{xy}{Q^4} + \frac{y+x}{y-x}log\frac{y}{x}
\eeq
To construct the mass squared matrix of the Higgs scalars 
we need to compute the quantities
\beq
M_{ab}^2=(\frac{\partial^2 V}{\partial \Phi_a\partial\Phi_b})_0
\eeq
where $\Phi_a$ (a=1-4) are defined by
\beq
\{\Phi_a\}= \{\phi_1,\phi_2, \psi_1,\psi_2\}
\eeq
and as already specified the subscript 0  means that we set 
$\phi_1=\phi_2=\psi_1=\psi_2=0$
after the evaluation of the mass matrix.
The tree and loop contributions to $M_{ab}^2$ are given by

\beq 
M_{ab}^2= M_{ab}^{2(0)}+ \Delta M_{ab}^2 
\eeq
where  $M_{ab}^{2(0)}$ are the contributions at the tree level and 
 $\Delta M_{ab}^2$ are the loop contributions where 
 
\beq
\Delta M_{ab}^2=
\frac{1}{32\pi^2}
Str(\frac{\partial M^2}{\partial \Phi_a}\frac{\partial M^2}{\partial\Phi_b}
log\frac{M^2}{Q^2}+M^2 \frac{\partial^2 M^2}{\partial \Phi_a\partial \Phi_b}
log\frac{M^2}{eQ^2})_0
\eeq
 where e=2.718.
Computation of the $4\times 4$ Higgs  mass matrix in the basis 
of Eq.(16) gives

\beq
\left(\matrix{M_Z^2c_{\beta}^2+M_A^2s_{\beta}^2+\Delta_{11} &
-(M_Z^2+M_A^2)s_{\beta}c_{\beta}+\Delta_{12} &\Delta_{13}s_{\beta}&\Delta_{13}
 c_{\beta}\cr
-(M_Z^2+M_A^2)s_{\beta}c_{\beta}+\Delta_{12} &
M_Z^2s_{\beta}^2+M_A^2c_{\beta}^2+\Delta_{22} & \Delta_{23} s_{\beta}
&\Delta_{23} c_{\beta}\cr
\Delta{13} s_{\beta} & \Delta_{23} s_{\beta}&(M_A^2+\Delta_{33})s_{\beta}^2 & 
(M_A^2+\Delta_{33})s_{\beta}c_{\beta}\cr
\Delta_{13} c_{\beta} &\Delta_{23} c_{\beta} &(M_A^2+\Delta_{33})s_{\beta}c_{\beta} & 
(M_A^2+\Delta_{33})c_{\beta}^2}\right)
\eeq
where $(c_{\beta}, s_{\beta})=(\cos\beta, \sin\beta)$.
In the above the explicit Q dependence has been absorbed in $m_A^2$
which is given by
\beqn
m_A^2=(\sin\beta\cos\beta)^{-1}(-m_3^2\cos\theta_H +\frac{1}{2}\beta_{h_t} 
|A_t||\mu|\cos\gamma_t f_1(m_{\tilde t_1}^2,m_{\tilde t_2}^2)\nonumber\\
+\frac{1}{2}\beta_{h_b} |A_b||\mu| \cos\gamma_b f_1(m_{\tilde b_1}^2,m_{\tilde b_2}^2) +\frac{g_2^2}{16\pi^2}|\tilde m_2||\mu| \cos\gamma_2 f_1(m_{\chi_1^+}^2, m_{\chi_2^+}^2))
 \eeqn
The first term in the second 
 brace on the right hand side of Eq.(20) is the tree term, 
while the second, the
third and the fourth terms come from the stop,sbottom and chargino
exchange contributions. We give now our computation of the $\Delta$'s.
For $\Delta_{ij}$ one has 
\beq
\Delta_{ij}=\Delta_{ij\tilde t}+\Delta_{ij\tilde b}+
\Delta_{ij\chi^+}
\eeq
where $\Delta_{ij\tilde t}$ is the contribution from the 
stop exchange in the loops, $\Delta_{ij\tilde b}$ is the 
contribution from the sbottom exchange in the loops and
$\Delta_{ij\chi^+}$ is the contribution from the chargino 
 sector in the loops. 
 $\Delta_{ij\tilde t}$ and  $\Delta_{ij\tilde b}$ are listed
in Appendix A. In the analysis of the chargino exchange we
shall approximate the chargino eigen values given by 
Eq.(11) by 

\beqn
M_{\chi_{1,2}^+}^2 \simeq M_W^2+ \frac{1}{2}
(|\tilde m_2|^2+|\mu|^2)\nonumber\\
\pm [\frac{1}{4}(|\tilde m_2|^2-|\mu|^2)^2 -M_W^2\cos 2\beta
(|\tilde m_2|^2-|\mu|^2)+ 
 2M_W^2 |\tilde m_2 \cos\beta+\mu^* \sin\beta|^2]^{\frac{1}{2}}
\eeqn
where we have ignored the term of O($M_W^4$) inside the
square root. This approximation leads  us to achieve
 an independence on Q of the chargino-W-$H^+$  exchange
correction and is similar to the approximation of dropping 
the D terms in the squark masses in the stop exchange correction
(see Appendix A). Below we list the result of our analysis of the
 $\Delta_{ij\chi^+}$ chargino exchange contributions. They are given by

\beqn
\Delta_{11\chi^+}=\frac{g_2^2}{8\pi^2} M_W^2 
\frac{((|\tilde m_2|^2+|\mu|^2)\cos\beta + 2|\tilde m_2||\mu|\sin\beta\cos\gamma_2)^2}{(m_{\chi_1^+}^2-m_{\chi_2^+}^2)^2}
f_2(m_{\chi_1^+}^2, m_{\chi_2^+}^2)\nonumber\\
-\frac{g_2^2}{4\pi^2} M_W^2
\frac{((|\tilde m_2|^2+|\mu|^2)\cos^2\beta + |\tilde m_2||\mu| \sin2\beta
\cos\gamma_2)}{(m_{\chi_1^+}^2- m_{\chi_2^+}^2)} 
ln(\frac{m_{\chi_1^+}^2}{ m_{\chi_2^+}^2})\nonumber\\
 -\frac{g_2^2}{16\pi^2} M_W^2\cos^2\beta
ln(\frac{m_{\chi_1^+}^4 m_{\chi_2^+}^4}{M_W^6M_{H^+}^2})
\eeqn
where 
\beq 
f_2(x,y)=-2+\frac{y+x}{y-x}ln\frac{y}{x}
\eeq

\beqn
\Delta_{22\chi^+}=\frac{g_2^2}{8\pi^2} M_W^2 
\frac{((|\tilde m_2|^2+|\mu|^2)\sin\beta + 2|\tilde m_2||\mu|\cos\beta\cos\gamma_2)^2}{(m_{\chi_1^+}^2-m_{\chi_2^+}^2)^2}
f_2(m_{\chi_1^+}^2, m_{\chi_2^+}^2)\nonumber\\
-\frac{g_2^2}{4\pi^2} M_W^2
\frac{((|\tilde m_2|^2+|\mu|^2)\sin^2\beta + |\tilde m_2||\mu| \sin2\beta
\cos\gamma_2)}{(m_{\chi_1^+}^2- m_{\chi_2^+}^2)} 
ln(\frac{m_{\chi_1^+}^2}{ m_{\chi_2^+}^2})\nonumber\\
 -\frac{g_2^2}{16\pi^2} M_W^2\sin^2\beta
ln(\frac{m_{\chi_1^+}^4 m_{\chi_2^+}^4}{M_W^6M_{H^+}^2}) 
\eeqn

\beqn
\Delta_{12\chi^+}=\frac{g_2^2}{8\pi^2} M_W^2 
[(|\tilde m_2|^2+|\mu|^2)\cos\beta + 2|\tilde m_2|
|\mu|\sin\beta\cos\gamma_2)]\nonumber\\
\frac{[(|\tilde m_2|^2+|\mu|^2)\sin\beta + 2|\tilde m_2||\mu|\cos\beta\cos\gamma_2)]}{(m_{\chi_1^+}^2-m_{\chi_2^+}^2)^2}
f_2(m_{\chi_1^+}^2, m_{\chi_2^+}^2)\nonumber\\
-\frac{g_2^2}{8\pi^2} M_W^2
\frac{((|\tilde m_2|^2+|\mu|^2)\sin 2\beta +2|\tilde m_2||\mu|
\cos\gamma_2)} { (m_{\chi_1^+}^2-m_{\chi_2^+}^2)  }
ln(\frac{m_{\chi_1^+}^2}{ m_{\chi_2^+}^2})\nonumber\\
 -\frac{g_2^2}{32\pi^2} M_W^2 \sin 2\beta
ln(\frac{m_{\chi_1^+}^4 m_{\chi_2^+}^4}{M_W^6M_{H^+}^2})
\eeqn

\beqn
\Delta_{13\chi^+}=\frac{g_2^2}{4\pi^2} M_W^2 
[(|\tilde m_2|^2+|\mu|^2)\cos\beta + 2|\tilde m_2|
|\mu|\sin\beta\cos\gamma_2)]\nonumber\\
\frac{|\tilde m_2||\mu|\sin\gamma_2}
{ (m_{\chi_1^+}^2-m_{\chi_2^+}^2)^2}
f_2(m_{\chi_1^+}^2, m_{\chi_2^+}^2)
-\frac{g_2^2}{4\pi^2} M_W^2
\frac{|\tilde m_2||\mu|\sin\gamma_2\cos\beta}
{  (m_{\chi_1^+}^2- m_{\chi_2^+}^2)}
ln(\frac{m_{\chi_1^+}^2}{ m_{\chi_2^+}^2})
\eeqn

\beqn
\Delta_{23\chi^+}=\frac{g_2^2}{4\pi^2} M_W^2 
|\tilde m_2||\mu|\sin\gamma_2\nonumber\\ 
\frac{[(|\tilde m_2|^2+|\mu|^2)\sin\beta + 2|\tilde m_2| |\mu| 
\cos\beta\cos\gamma_2]}
{(m_{\chi_1^+}^2-m_{\chi_2^+}^2)^2}
f_2(m_{\chi_1^+}^2, m_{\chi_2^+}^2)\nonumber\\
-\frac{g_2^2}{4\pi^2} M_W^2
\frac{|\tilde m_2||\mu|\sin\gamma_2 \sin\beta}
{(m_{\chi_1^+}^2- m_{\chi_2^+}^2)} 
ln(\frac{m_{\chi_1^+}^2}{ m_{\chi_2^+}^2})
\eeqn

\beq
\Delta_{33\chi^+}=\frac{g_2^2}{2\pi^2} 
\frac{M_W^2 |\tilde m_2|^2|\mu|^2\sin^2\gamma_2} 
{(m_{\chi_1^+}^2- m_{\chi_2^+}^2)^2}
f_2(m_{\chi_1^+}^2, m_{\chi_2^+}^2)
\eeq
We note that all the $\Delta_{ij\chi^+}$ have no explicit 
Q dependence. Inclusion of the W and the $H^+$ exchange along 
with the chargino exchange was necessary to achieve the Q 
independence. 
We further note that unlike the third generation contributions where
one needs to worry about the possibility of  significant 
QCD corrections, the chargino exchange is purely electro-weak
in nature and thus largely free of such corrections. 
Eqs.(20-29) constitute the main new theoretical computations
in this paper. Using these equations and the results of the 
analysis of Appendix A one finds $\Delta_{ij}$ of Eq.(21)
and thus computes the  matrix of Eq.(19). One may reduce the 
$4\times 4$ matrix of Eq.(19) by introducing a 
 new basis $\{ \phi_1,\phi_2,\psi_{1D}, \psi_{2D}\}$
where $\psi_{1D},\psi_{2D}$ are defined by 
\beqn
\psi_{1D}=\sin\beta \psi_1+ \cos\beta \psi_2\nonumber\\
\psi_{2D}=-\cos\beta \psi_1+\sin\beta \psi_2
\eeqn
In this basis the field $\psi_{2D}$ decouples from the other three
fields. $\psi_{2D}$ is a zero mass state and is the Goldstone field. The
  Higgs $(mass)^2$ matrix $M^2_{Higgs}$ of the remaining three 
fields is given
  by
\beq
M^2_{Higgs}=
\left(\matrix{M_Z^2c_{\beta}^2+M_A^2s_{\beta}^2+\Delta_{11} &
-(M_Z^2+M_A^2)s_{\beta}c_{\beta}+\Delta_{12} &\Delta_{13}\cr
-(M_Z^2+M_A^2)s_{\beta}c_{\beta}+\Delta_{12} &
M_Z^2s_{\beta}^2+M_A^2c_{\beta}^2+\Delta_{22} & \Delta_{23} \cr
\Delta_{13} & \Delta_{23} &(M_A^2+\Delta_{33})}\right)
\eeq
We note that in principle it is possible that due to cancellations
between the stop and the chargino contributions to Eq.(12)
that $\theta_H$ vanishes or becomes very small. However, even in
this case  the 
mixings between the CP even Higgs sector and the CP odd Higgs
sector can still occur because the parameters that determine
this mixings are $\gamma_t$, $\gamma_b$ and $\gamma_2$ and 
not $\theta_H$. $\gamma_t$ gets directly related to $\theta_H$
when contributions to $\theta_H$ other than the stop exhanges
are ignored.

We can obtain an approximation to the chargino corrections
to the Higgs masses using a perturbation expansion. 
We order the eigen values so that in the limit of 
no mixing between the CP even and the CP odd states one has
$(m_{H_1}, m_{H_2}, m_{H_3})$ $\rightarrow$
$(m_H, m_h, m_A)$. 
Defining $m_h=m_h^0+(\Delta m_h)_{\chi^+}$ 
where $m_h^0$ is the lightest Higgs mass without the
chargino loop contribution, and $(\Delta m_h)_{\chi^+}$ is 
the correction due to the chargino exchange loops, and
with $(\Delta  m_H)_{\chi^+}$ and  $(\Delta m_A)_{\chi^+}$
similarly defined, one finds 
\beqn
 (\Delta m_H)_{\chi^+} =(2m_H^0)^{-1}
(\Delta_{11\chi^+  } \cos^2\alpha 
+\Delta_{22\chi^+} \sin^2\alpha + \Delta_{12} \sin2\alpha)\nonumber\\
  (\Delta m_h)_{\chi^+}=(2m_h^0)^{-1}
(\Delta_{11\chi^+ } \sin^2\alpha 
+\Delta_{22\chi^+ } \cos^2\alpha - \Delta_{12\chi^+ } \sin2\alpha)\nonumber\\
  (\Delta m_A)_{\chi^+} =(2m_A^0)^{-1} \Delta_{33\chi^+ }
\eeqn
where 
\beqn 
\cos 2\alpha\simeq\frac{M_{11}^2-M_{22}^2}{\sqrt{(trM^2)^2-4(detM^2)}}\nonumber\\
\sin 2\alpha\simeq\frac{2M_{12}^2}{\sqrt{(trM^2)^2-4(detM^2)}}
\eeqn
where the matrix $M^2$ is the $2\times 2$ matrix in the upper
left hand corner of Eq.(31), i.e.,
\beq
(M^2)=\left(\matrix{M_{11}^2 & M_{12}^2\cr
M_{21}^2& M_{22}^2}\right)
=
\left(\matrix{M_Z^2c_{\beta}^2+M_A^2s_{\beta}^2+\Delta_{11} &
-(M_Z^2+M_A^2)s_{\beta}c_{\beta}+\Delta_{12} \cr
-(M_Z^2+M_A^2)s_{\beta}c_{\beta}+\Delta_{12} &
M_Z^2s_{\beta}^2+M_A^2c_{\beta}^2+\Delta_{22}}\right)
\eeq
Numerically the approximation of Eq.(32) turns out to be 
accurate to within a 
few percent compared to the exact results obtained from 
diagonalization of the $3\times 3$ matrix of Eq.(31).

\section{Size of chargino sector loop contributions}
We discuss now the numerical size of the chargino sector exchange
contributions. The current lower limits on the light Higgs masses
correspond to $m_h>88.3$ GeV for the light CP even Higgs 
and $m_A>88.4$ GeV for the CP odd Higgs\cite{lep}.
In our analysis
we shall examine the part of the MSSM parameter space where these
limits are obeyed although it should be kept in mind that the
analysis leading to these limits included no CP violating 
effects.\cite{carena}.
 Since the general parameter  space of MSSM is 
rather large, we shall limit ourselves to a more constrained set
for the purpose of this numerical study. We shall use for our parameter space 
the set $m_0, m_{\frac{1}{2}}$, $m_A$, $|A_0|$, $\tan\beta$,
$\theta_{\mu}$, $\alpha_{A_0}$, $\xi_1$, $\xi_2$ and $\xi_3$.
 The parameter $\mu$ is determined via radiative breaking of the
electro-weak symmetry.
The other sparticle masses  are obtained from this set using the
 renormalization group equations evolving the GUT parameters from the 
 GUT scale
down to the electro-weak scale. However, this is  only a convenience and 
in general one can use the general MSSM parameter space to 
test the size of the corrections computed here.
As discussed in Ref.\cite{in1,in2,in3} there exists a very
significant part in the MSSM parameter  space  where  the 
EMD constraints are satisfied with large phases. For the 
purposes  of this analysis we shall assume that this is the
case and not revisit the problem of imposing the edms constraints.
 We note that in the general analysis using the  MSSM parameter space
the edms depend on 10 separate phases and thus
alternately one may view the phases we vary as unconstrained while
other phases which do not enter in the analysis as restricted by 
the edm constraints. 

  In the numerical analysis we  consider
the contributions from the stop, the sbottom as well as 
from the chargino exchange. As can be seen from Eqs.(23)-(29) 
 the chargino contribution depends 
on the combination  $\xi_2+\theta_{\mu}$.
However, the stop contribution depends on the combination
 $\alpha_{A_t}+\theta_{\mu}$, and the sbottom contribution depends on the
combination  $\alpha_{A_b}+\theta_{\mu}$ (see Appendix A).
 Clearly the total contribution is thus a function of three independent
phases. Specifically the total contribution will in general have
a different  dependence on $\xi_2$ and  $\theta_{\mu}$ when the
other parameters are kept fixed. Now firstly, we study
the variation of the total contribution on $\xi_2$ since this is a  new
phase that does not appear in the stop and sbottom contributions which
have been discussed in the previous literature.
 Thus as we vary $\xi_2$ the stop and the sbottom
contribution remains constant while the chargino contribution alone
varies. However, the variation of $\theta_{\mu}$ reveals a different
dependence since this time all contributions, ie., the stop, the 
sbottom as well as the chargino contribution individually vary 
as we vary $\theta_{\mu}$. We discuss now the numerical analysis 
in detail.

Using the constrained parameter space described above we plot in 
Fig.1 the quantity  $\Delta_{13}$
as a function of the CP phase  $\xi_2$. The  $\Delta_{13}$ plots 
exhibited in Fig.1  contain
the stop, the sbottom and the chargino sector contributions 
while the horizontal lines exhibit $\Delta_{13}$ without the 
inclusion of the chargino sector contribution.
The analysis shows that the chargino sector contribution to $\Delta_{13}$
is comparable to the stop exchange  contribution. Further one finds
 that the chargino sector contribution 
 can be either positive or negative relative to the stop and sbottom
 sector
 contribution. Thus the chargino sector contribution can
 constructively interfere with the stop and sbottom sector contribution 
 enhancing the CP even and CP odd mixing by as much as  a factor 
 of two. However, in other regions of the parameter space it can
 produce a negative interference reducing significantly the mixing of the
   CP even and CP odd Higgs sectors. 
 A similar analysis holds for $\Delta_{23}$ and 
 in Fig.2 we give a plot of $\Delta_{23}$, with and without 
 the contribution from the chargino sector,  as a function of $\xi_2$. 
  One finds again that the chargino sector
  makes a large contributions  to   $\Delta_{23}$.
 Further, as for the  case of $\Delta_{13}$, the chargino sector
 contributions can  either constructively or destructively 
 interfere with the stop and sbottom sector contribution and thus the 
 chargino sector 
 contribution can either 
 enhance or reduce the size of $\Delta_{23}$.

  In Fig.3 a plot of the percentage of the
 CP even component $\phi_1$ of $H_1$ (upper sets) and the CP odd 
 component $\psi_{1D}$ of $H_1$ (lower sets) including the stop, 
 the sbottom and the chargino sector 
 contributions is given as a function of $\xi_2$ while the 
 horizontal lines give the plots when the chargino sector  
 contribution is omitted. (The $\phi_2$ component is negligible 
 and is not exhibited.)  A comparison of the plots with and without
 the chargino sector contribution shows that the chargino sector 
 makes a large relative contribution to the $\phi_1$ and the $\psi_{1D}$ 
 components
 and further that this  
 contribution  can either constructively or destructively interfere 
 with the contribution coming from the stop and sbottom sector
 contribution. 
  We note that for the inputs of Fig.3 the mixings between
 the CP even and the CP odd components are essentially maximal.
 This phenomenon is a consequence of large $\tan\beta$ and
 we will study this in greater depth when we discuss the analysis
 of Fig.4. 
  A similar analysis for  
 $H_2$ yields a much smaller effect, i.e., less than a percent or
 so for this  case where $H_2$ is the eigen state which limits
 to the lightest  CP even Higgs state in the case when one
 ignores the  mixing between the CP even and the CP odd states. 
 Thus one concludes that the
 lightest Higgs state develops a negligible CP odd  component
 as a consequence  of mixing and remains essentially a CP even state.
 A similar conclusion was arrived at in previous 
 analyses\cite{demir} without the inclusion of the
 chargino sector contribution. The analysis of 
 $H_3$ parallels the analysis of $H_1$ except that the
 CP even and the CP odd components reverse their roles. Thus one can
 easily obtain the percentages of $\phi_1$ and $\psi_{1D}$ components 
 in $H_3$ from Fig.3 by interchanging $\phi_1$ and $\psi_{1D}$. Again
 one finds that the $\phi_2$ component of $H_3$ is small for the 
 input of Fig.3. 
 
 We discuss now the $\tan\beta$ dependence of the mixing  
 between the CP even and the CP odd sectors. 
  We illustrate this dependence in Fig.4 where the CP odd
  component $\psi_{1D}$ of $H_1$ is plotted as a function of $\xi_2$
  for values of $\tan\beta$ ranging from 5 to 40. One finds that
  for  $\tan\beta\leq 10$ the CP odd component of $H_1$ is 
  less  than a fraction of a percent. The fraction of the CP odd
  component grows to the level of 
  a few percent for values of  $\tan\beta$ in the range 15-20.
  This trend continues and  one finds large mixings as $\tan\beta$
  gets large, ie., in the neighborhood of 25 or larger. 
  The theoretical reason for this strong dependence of the mixings
  on $\tan\beta$ can be easily understood. Thus as $\tan\beta$ 
  becomes large  $\cos\beta$ becomes vanishingly small, and from
  Eq.(31) one finds that the two heavier Higgs eigen masses 
  become essentially degenerate. This degeneracy of masses  implies
  that the mixings are no longer suppressed by the factor  
  $\Delta_{ij}/M_A^2$ etc but rather it is the ratio of the
  $\Delta's$ themselves that determines the mixings. Consequently
  in the region of large $\tan\beta$ the mixings between the CP
  even and the CP odd sectors become large. In the analysis  presented
  so far we have investigated the dependence of the mixings of the
  CP even and the CP odd sector on $\xi_2$ which is the phase of the 
  $SU(2)$ gaugino mass $\tilde m_2$. 
   One also expects a significant dependence of
  the mixings on the other phases. As an illustration in Fig.5
  we give an analysis of the mixings as a function of $\theta_{\mu}$.
  Here, as in Fig.3, we  plot the CP even and the CP odd components of
  $H_1$, i.e., of $\phi_1$ and of $\psi_{1D}$ but now as a function of
  $\theta_{\mu}$ for the inputs given in the figure caption. 
  The dashed curves are for the case without the chargino sector contribution
  while
  the solid curves are with the chargino sector contribution. 
  Again one finds that
  the chargino sector makes a significant contribution relative
  to the stop and sbottom sector contribution.   
  
 Finally, we discuss the contribution of the
 chargino sector to the lightest Higgs mass. One finds 
 that the chargino sector contribution is typically
 of order 1-2 GeV and is negative. Some typical examples
 of the sizes of the $\chi^+-W-H^+$ contribution are given in
 Table 1. The effect of CP phases on the $\chi^+-W-H^+$  correction
 to the Higgs masses is typically small, i.e., the variation in the
 corrections is a few percent at best.   
 The precision analyses
 of the Higgs masses including radiative corrections
 from the stop and sbottom sector exchanges and including 
 leading order  corrections from two loop corrections
 and other refinements purport to achieve an 
 accuracy of 1-2 GeV in the prediction of the lightest Higgs boson mass.
 Since the chargino sector contribution with or without 
 CP violating  effects lies in this range it appears reasonable  
 to include this  correction in the precision prediction of 
 the lightest Higg boson mass. The chargino sector corrections to the 
 mass eigen values of the other two ($H_1, H_3$) Higgs bosons
 is significantly smaller and can be safely neglected.

\begin{center} \begin{tabular}{|c|c|c|}
\multicolumn{3}{c}{Table~1:  } \\
\hline
  $\tan\beta$& $m_h$ without $\chi^+-W-H^+$ & $m_h$ with $\chi^+-W-H^+$\\
 \hline
 5 & 116.82 & 115.42\\
 \hline
 10 &121.76 &120.45\\
 \hline
 15 & 122.70 & 121.41\\
 \hline
 20 & 123.0 & 121.71 \\
 \hline
 25 & 123.11 & 121.84 \\
 \hline
 30 & 123.16 & 121.88 \\
 \hline
\end{tabular}\\
\noindent
Table caption: Input parameters are $m_0=500$, $m_{\frac{1}{2}}=400$,
$m_A=200$, $|A_0|=1000$, $\theta_{\mu}=0.5$, $\alpha_{A_0}=0.5$,
$\xi_1=0.4$, $\xi_2=0.5$, $\xi_3=0.6$. All masses are in GeV and
all angles are in radians.
\end{center}

   There are several consequences of the CP even and the CP odd Higgs
mixings implied by large phases. Some of these have been 
discussed in Refs.\cite{demir,carena}. One consequence is 
  the effect on the quark and on the lepton couplings with the Higgs, i.e.,
the couplings $\bar q qH_i$ and $\bar l lH_i$ (i=1,2,3). 
These modifications affect the phenomenology for Higgs searches
at colliders. We point out here that the vertices involving
the couplings of the Higgs with the charginos ($\chi_a^+$, a=1,2)
and the 
neutralinos ($\chi_n, n=1-4$) are also affected, i.e., the 
chargino Higgs couplings
$\bar\chi_a^+\chi_b^+H_i$ (a,b=1,2), and the neutralino Higgs couplings
$\bar\chi_n\chi_mH_i$ (n,m=1-4). 
 Specifically, the couplings of the lightest
neutralino ($\chi_1$) to the  Higgs will depend on the
parameters which mix the CP even and the CP odd Higgs sector 
and will affect
dark matter analyses. Thus  the  neutralino relic density analysis
 which involves  the process
$ \chi_1+\chi_1$ $\rightarrow$ $\bar ff$ etc,
with  the Higgs poles apprearing in the direct channel,
will be affected. We expect these effects to arise 
from the couplings of the $H_1$ and $H_3$ and expect them to  give
significant effects only for large values of $\tan\beta$,
i.e., $\tan\beta >20$ where the mixing effects become significant. 
Similarly the analysis of the direct detection 
of dark matter which involves the scattering  process 
$\chi_1+ q \rightarrow \chi_1+q$, with the Higgs poles
 entering in the cross channel, will be affected.
      A detailed  discussion of these
phenomena is outside the scope of this paper.

\section{Conclusions}
In this paper we have analysed the effects of the $\chi^+$, the W and 
the $H^+$
exchange contributions to the Higgs boson masses and mixings in 
the presence of  large CP violating effects. We find that this sector 
makes a large contribution to the mixing between the CP even and the
CP odd Higgs states and  in certain
parts of the parameter space the  mixings generated by the chargino
sector  may dominate the mixings generated by the stop and
sbottom sector exchanges. 
We also find that in terms of sizes 
 the chargino sector contributions are significantly larger
 than the sbottom exchange
corrections which have been included in previous analyses.
The size of the mixing effects are seen to depend sharply on 
the value of $\tan\beta$ with the mixing effects becoming large
as $\tan\beta$ gets large and for values of $\tan\beta$ larger than
30 the mixings between the CP even and the CP odd sector 
become maximal. These mixings have
important implication for Higgs phenomenology at colliders.
We have also analysed the effects of the chargino sector  
contribution on the lightest Higgs mass. We find that the chargino sector
 contribution to the lightest Higgs boson mass 
 lies  in the range
of 1-2 GeV. This effect is relevant in the precision predictions
of the lightest Higgs boson mass.
 Further, we find that typically the chargino sector contribution is negative 
 and lowers the lightest Higgs boson mass and leads to a slight 
 worsening of the fine tuning problem already implied by the 
 non observation of the Higgs boson at LEP thus far\cite{ccn}.
 A similar analysis can be
carried out for the neutralino sector  contribution to 
the Higgs boson masses and mixings.  This sector is significantly
more  difficult and requires new techniques for its analysis  because 
the neutralino mass matrix is $4\times 4$  and cannot be diagonalized 
with the same ease as the chargino or the squark sector can be. 
This analysis is underway and will be reported in a separate communication.

\noindent
{\bf Acknowledgments}\\ 
P.N. thanks the Physics Institute at the University of Bonn,
 where part of the work was done, for hospitality and  acknowledges 
support from an Alexander von Humboldt award.
 This research was also supported in part by NSF grant PHY-9901057\\

\section{Appendix A: Stop and sbottom contributions}
For completeness we give here an analysis of the one loop 
contributions from the stop and 
sbottom sectors with inclusion of CP violating effects. 
The stop $(mass)^2$ matrix is given by

\beq
M^2_{\tilde t}=
\left(\matrix{M_Q^2+h_t^2|H_2^0|^2+ \frac{(g_2^2-g_1^2/3)}{4}
(|H_1^0|^2-|H_2^0|^2) & h_t(A_t^* H_2^{0*}-\mu H_1^0)\cr
  h_t(A_t H_2^{0}-\mu^* H_1^{0*}) & M_U^2+h_t^2|H_2^0|^2+ \frac{g_1^2}{3}
(|H_1^0|^2-|H_2^0|^2)}\right)
\eeq
where $A_t=|A_t|e^{i\alpha_{A_t}}$. 
The contribution to the one loop effective potential from
the stop and top exchanges is given by

\beq
\Delta V(\tilde t, t)=\frac{1}{64\pi^2}
(\sum_{a=1,2}6M_{\tilde t_a}^4(log \frac{M_{\tilde t_a}^2}{Q^2}-\frac{3}{2})
-12 m_t^4 (log\frac{m_t^2}{Q^2}-\frac{3}{2}))
\eeq
Using the above potential our analysis for $\Delta_{ij\tilde t}$ gives

\beq
\Delta_{11\tilde t}=-2\beta_{h_t}m_t^2 |\mu|^2 
\frac{(|A_t|\cos\gamma_t -|\mu|cot\beta)^2}
{(m_{\tilde t_1}^2-m_{\tilde t_2}^2)^2} 
f_2(m_{\tilde t_1}^2, m_{\tilde t_2}^2)
\eeq

\beqn
\Delta_{22\tilde t}=-2\beta_{h_t} m_t^2 
\frac{|A_t|^2[|A_t| -|\mu|cot\beta\cos\gamma_t]^2}
{(m_{\tilde t_1}^2-m_{\tilde t_2}^2)^2} 
f_2(m_{\tilde t_1}^2, m_{\tilde t_2}^2)\nonumber\\
+2\beta_{h_t} m_t^2 
ln(\frac{m_{\tilde t_1}^2 m_{\tilde t_2}^2}{m_t^4})
+ 4\beta_{h_t}m_t^2
\frac{|A_t|[|A_t| -|\mu|cot\beta\cos\gamma_t]}
{(m_{\tilde t_1}^2-m_{\tilde t_2}^2)} 
ln(\frac{m_{\tilde t_1}^2}{m_{\tilde t_2}^2})
\eeqn

\beqn
\Delta_{12\tilde t}=
-2\beta_{h_t} m_t^2 
\frac{|\mu|[|A_t|\cos\gamma_t -|\mu|cot\beta]}
{(m_{\tilde t_1}^2-m_{\tilde t_2}^2)}
ln(\frac{m_{\tilde t_1}^2}{ m_{\tilde t_2}^2})\nonumber\\
+ 2\beta_{h_t}m_t^2
\frac{|\mu||A_t|[|A_t|\cos\gamma_t -|\mu|cot\beta]
[|A_t| -|\mu|cot\beta\cos\gamma_t]}
{(m_{\tilde t_1}^2-m_{\tilde t_2}^2)^2} 
f_2(m_{\tilde t_1}^2, m_{\tilde t_2}^2)
\eeqn

\beq
\Delta_{13\tilde t}=
-2\beta_{h_t} m_t^2 
\frac{|\mu|^2|A_t|\sin\gamma_t [|\mu|cot\beta -|A_t|\cos\gamma_t ]}
{\sin\beta(m_{\tilde t_1}^2-m_{\tilde t_2}^2)^2}
f_2(m_{\tilde t_1}^2, m_{\tilde t_2}^2)
\eeq

\beqn
\Delta_{23\tilde t}=
-2\beta_{h_t} m_t^2 |\mu||A_t|^2
\frac{\sin\gamma_t (|A_t| - |\mu|cot\beta \cos\gamma_t)}
{\sin\beta(m_{\tilde t_1}^2-m_{\tilde t_2}^2)^2}
f_2(m_{\tilde t_1}^2, m_{\tilde t_2}^2)\nonumber\\
+2\beta_{h_t}\frac{ m_t^2 |\mu||A_t|\sin\gamma_t}
{\sin\beta(m_{\tilde t_1}^2-m_{\tilde t_2}^2)}
ln(\frac{m_{\tilde t_1}^2}{ m_{\tilde t_2}^2})
\eeqn

\beq
\Delta_{33\tilde t}=
-2\beta_{h_t} \frac{m_t^2|\mu|^2|A_t|^2\sin^2\gamma_t}
{\sin^2\beta(m_{\tilde t_1}^2-m_{\tilde t_2}^2)^2}
f_2(m_{\tilde t_1}^2, m_{\tilde t_2}^2)
\eeq
The expressions of our  $\Delta_{11}$, $\Delta_{22}$,
$\Delta_{13}$, $\Delta_{23}$ and $\Delta_{33}$ agree with
those of previous authors. However, there is a difference between our
 $\Delta_{12}$ and the $\Delta_{12}$ of Ref.\cite{demir} in the
presence of phases although the two expressions agree in the limit 
when there are no phases.

We discuss next our computation for the sbottom sector.
The sbottom mass $(mass)^2$ matrix is given by 

\beq
M^2_{\tilde b}=
\left(\matrix{M_Q^2+h_b^2|H_1^0|^2- \frac{(g_2^2+g_1^2/3)}{4}
(|H_1^0|^2-|H_2^0|^2) & h_b(A_b^* H_1^{0*}-\mu H_2^0)\cr
  h_b(A_b H_1^{0}-\mu^* H_2^{0*}) & M_D^2+h_b^2|H_1^0|^2-\frac{g_1^2}{6}
(|H_1^0|^2-|H_2^0|^2)}\right)
\eeq
where $A_b=|A_b|e^{i\alpha_{A_b}}$. 
The contribution to the one loop effective potential from
the sbottom and b exchanges is given by

\beq
\Delta V(\tilde b, b)=\frac{1}{64\pi^2}
(\sum_{a=1,2} 6 M_{\tilde b_a}^4(log\frac{M_{\tilde b_a}^2}{Q^2}-\frac{3}{2})
-12 m_b^4 (log\frac{m_b^2}{Q^2}-\frac{3}{2}))
\eeq
Our computation of $\Delta_{ij\tilde b}$ yields
 
\beqn
\Delta_{11\tilde b}=-2\beta_{h_b} m_b^2 
\frac{|A_b|^2[|A_b| -|\mu|\tan\beta\cos\gamma_b]^2}
{(m_{\tilde b_1}^2-m_{\tilde b_2}^2)^2} 
f_2(m_{\tilde b_1}^2, m_{\tilde b_2}^2)\nonumber\\
+2\beta_{h_b} m_b^2 
ln(\frac{m_{\tilde b_1}^2 m_{\tilde b_2}^2}{m_b^4})
+ 4\beta_{h_b}m_b^2
\frac{|A_b|[|A_b| -|\mu|\tan\beta\cos\gamma_b]}
{(m_{\tilde b_1}^2-m_{\tilde b_2}^2)} 
ln(\frac{m_{\tilde b_1}^2}{m_{\tilde b_2}^2})
\eeqn

\beq
\Delta_{22\tilde b}=-2\beta_{h_b}m_b^2 |\mu|^2 
\frac{(|\mu|\tan\beta-|A_b|\cos\gamma_b)^2}
{(m_{\tilde b_1}^2-m_{\tilde b_2}^2)^2} 
f_2(m_{\tilde b_1}^2, m_{\tilde b_2}^2)
\eeq

\beqn
\Delta_{12\tilde b}=-2\beta_{h_b}m_b^2 |\mu| 
\frac{(|A_b|\cos\gamma_b -|\mu|\tan\beta)}
{(m_{\tilde b_1}^2-m_{\tilde b_2}^2)}
ln(\frac{m_{\tilde b_1}^2}{ m_{\tilde b_2}^2})\nonumber\\
+ 2\beta_{h_b}m_b^2
\frac{|\mu||A_b|[|A_b|\cos\gamma_b -|\mu|\tan\beta]
[|A_b| -|\mu|\tan\beta\cos\gamma_b]}
{(m_{\tilde b_1}^2-m_{\tilde b_2}^2)^2} 
f_2(m_{\tilde b_1}^2, m_{\tilde b_2}^2)
\eeqn

\beqn
\Delta_{13\tilde b}=-2\beta_{h_b}m_b^2 |\mu||A_b|^2 \sin\gamma_b 
\frac{(|A_b| -|\mu|\tan\beta\cos\gamma_b)}
{\cos\beta(m_{\tilde b_1}^2-m_{\tilde b_2}^2)^2}
f_2(m_{\tilde b_1}^2, m_{\tilde b_2}^2)\nonumber\\
+ 2\beta_{h_b}m_b^2
\frac{|\mu||A_b| \sin\gamma_b}
{\cos\beta(m_{\tilde b_1}^2-m_{\tilde b_2}^2)}
ln(\frac{m_{\tilde b_1}^2}{ m_{\tilde b_2}^2})
\eeqn

\beq
\Delta_{23\tilde b}=-2\beta_{h_b}m_b^2 |\mu|^2|A_b|\sin\gamma_b 
\frac{(|\mu|\tan\beta-|A_b|\cos\gamma_b)}
{\cos\beta(m_{\tilde b_1}^2-m_{\tilde b_2}^2)^2}
f_2(m_{\tilde b_1}^2, m_{\tilde b_2}^2)
\eeq

\beq
\Delta_{33\tilde b}=
-2\beta_{h_b} \frac{m_b^2|\mu|^2|A_b|^2\sin^2\gamma_b}
{\cos^2\beta(m_{\tilde b_1}^2-m_{\tilde b_2}^2)^2}
f_2(m_{\tilde b_1}^2, m_{\tilde b_2}^2)
\eeq
In the above analysis we have ignored the D terms of the squark 
(mass)$^2$ matrices to gain approximate independence of the
renormalization scale Q as in the analysis of Ref.\cite{demir,carena}. \\

\noindent
{\bf Figure Captions}\\
Fig.1: Plot of $\Delta_{13}$ 
including the stop, sbottom and chargino sector contributions 
vs $\xi_2$. The common input for both the top and the bottom
curves is $m_0=500$, $m_{\frac{1}{2}}=400$,
$M_A=200$, $|A_0|=1000$, $\alpha_0=0.5$,
$\xi_1=0.4$ and $\xi_3=0.6$ where all masses are in GeV and all
angles are in radians. The dashed  curve is the case 
$\tan\beta=30$, $\theta_{\mu}=1$  and the dashed  horizontal line is without
the inclusion of the chargino sector contribution. The
corresponding  solid curve and the solid horizontal line are
for the same input except that $\tan\beta=40$ and  $\theta_{\mu}=0.5$.\\

\noindent
Fig.2: Plot of $\Delta_{23}$ 
including the stop, sbottom and chargino sector contributions 
vs $\xi_2$ for the same input as in Fig.1. The dashed  and 
solid curves have the same meaning as in Fig.1 and the horizontal 
lines are the plots without inclusion of the chargino sector contribution 
also as in Fig.1.\\

\noindent
Fig.3: Plot of the CP even component $\phi_1$ of $H_1$ (upper 
curves)  and the CP odd component $\psi_{1D}$ of $H_1$ (lower curves)
including the stop, sbottom and chargino sector contributions 
  as a function of $\xi_2$  for the same inputs as in Fig.1. 
  The dashed  curves are for the case $\tan\beta=30$, $\theta_{\mu}=1$,
  and the solid curves are for the case $\tan\beta=40$ 
    $\theta_{\mu}=0.5$, and the  
  corresponding horizontal lines are for the cases when the chargino 
  sector contributions are neglected. \\

\noindent
Fig.4: Plot of the modulus square of the CP odd  component in 
$H_1$ as a function of $\xi_2$ for various values of $\tan\beta$
with all the other parameters being the same as in Fig.3 with 
$\theta_{\mu}=0.5$.  The 
values of $\tan\beta$  from bottom up are 5,10,15,20,30,40. \\

\noindent
Fig.5:  Plot of the CP even component $\phi_1$ of $H_1$ (upper 
curves)  and the CP odd component $\psi_{1D}$ of $H_1$ (lower curves)
as a function of $\theta_{\mu}$ including the stop, sbottom and chargino 
sector contributions (solid) and without inclusion of chargino 
contributions (dashed) for the following input:
$m_0=500$, $m_{\frac{1}{2}}=400$, $m_A=300$, $A_0=1000$,
$\tan\beta=30$, $\alpha_{A_0}=-0.4$, $\xi_1=0.4$, 
$\xi_2=0.5$, $\xi_3=0.6$, where all masses are in GeV and 
all angles are in radians.\\

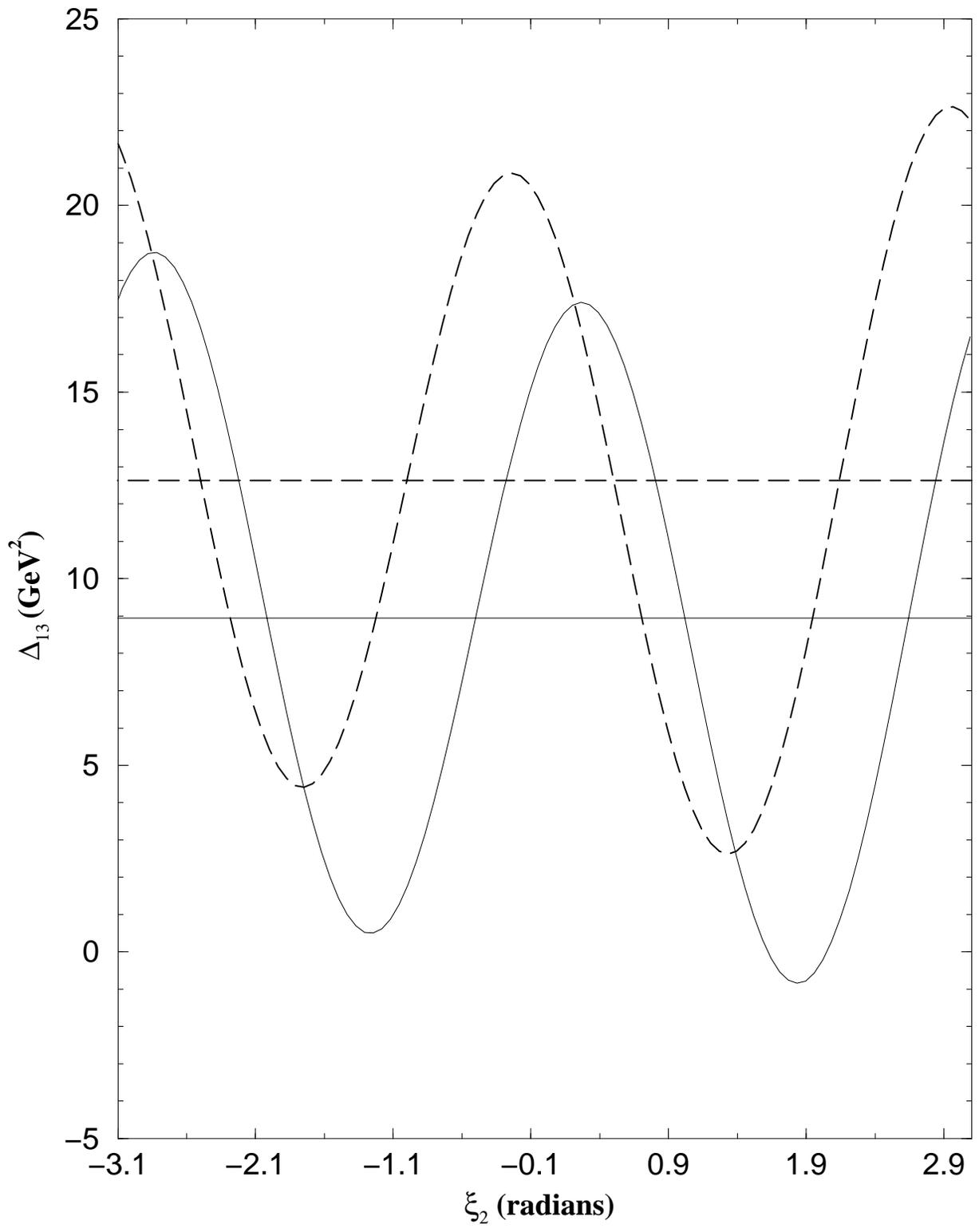
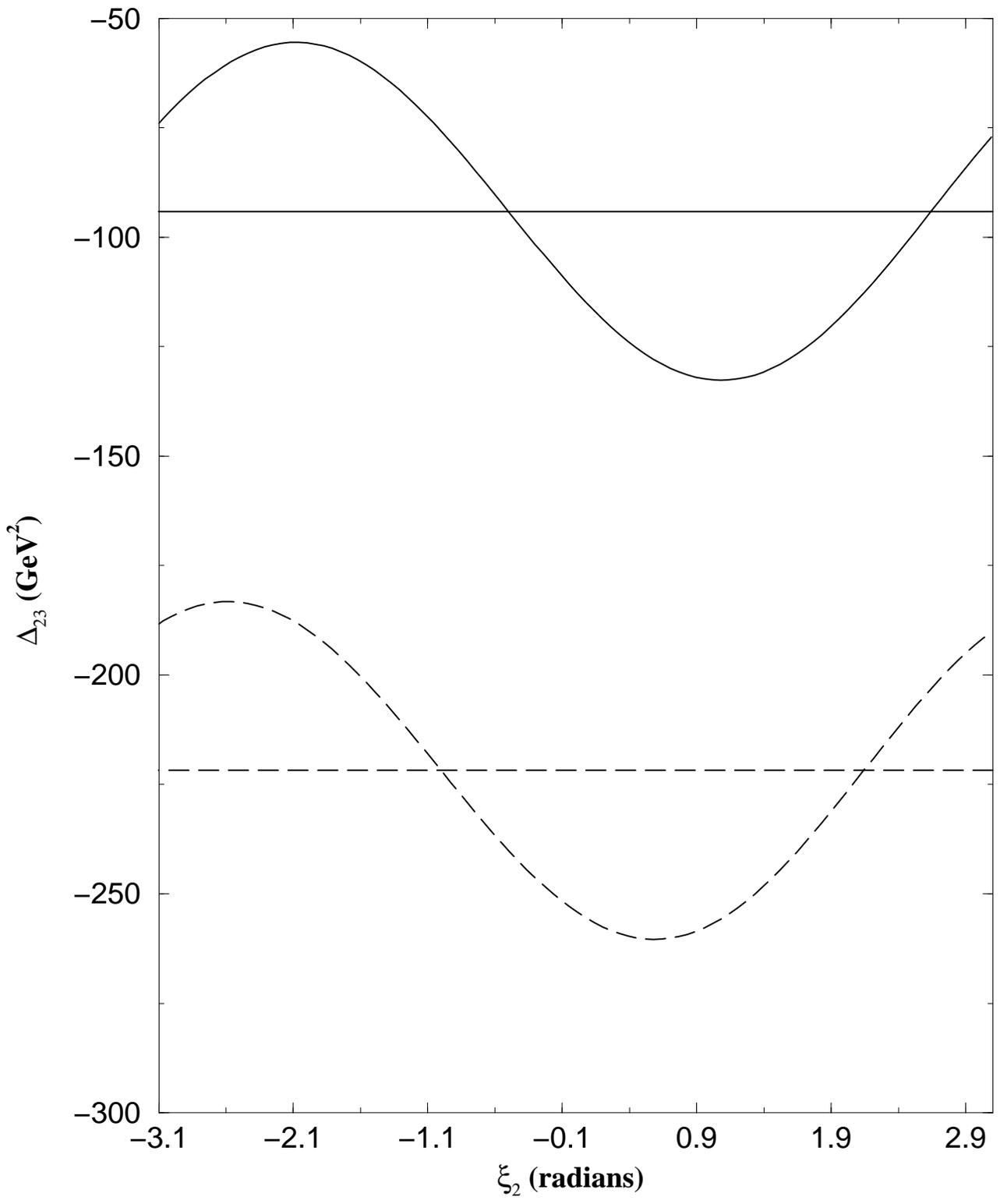
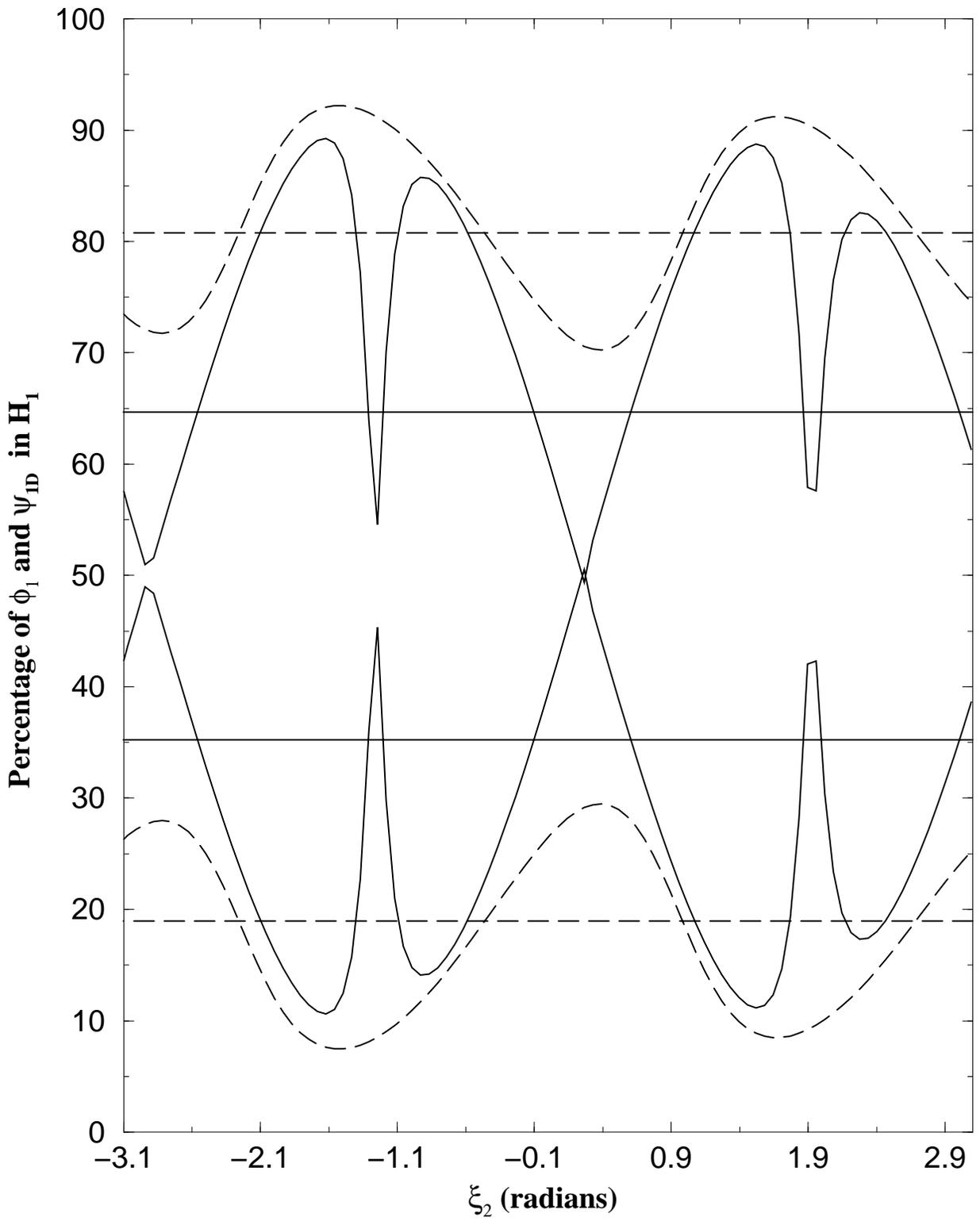
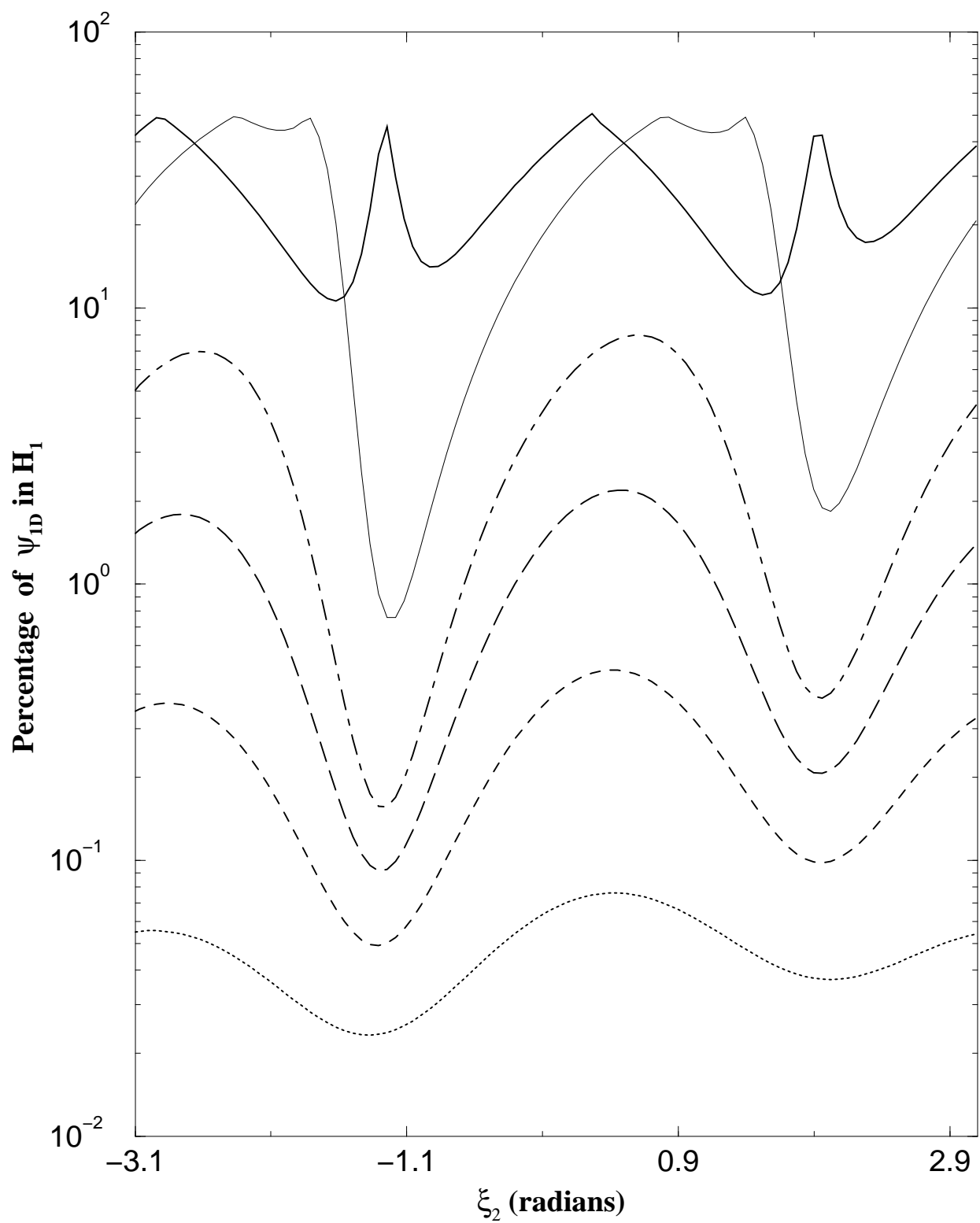
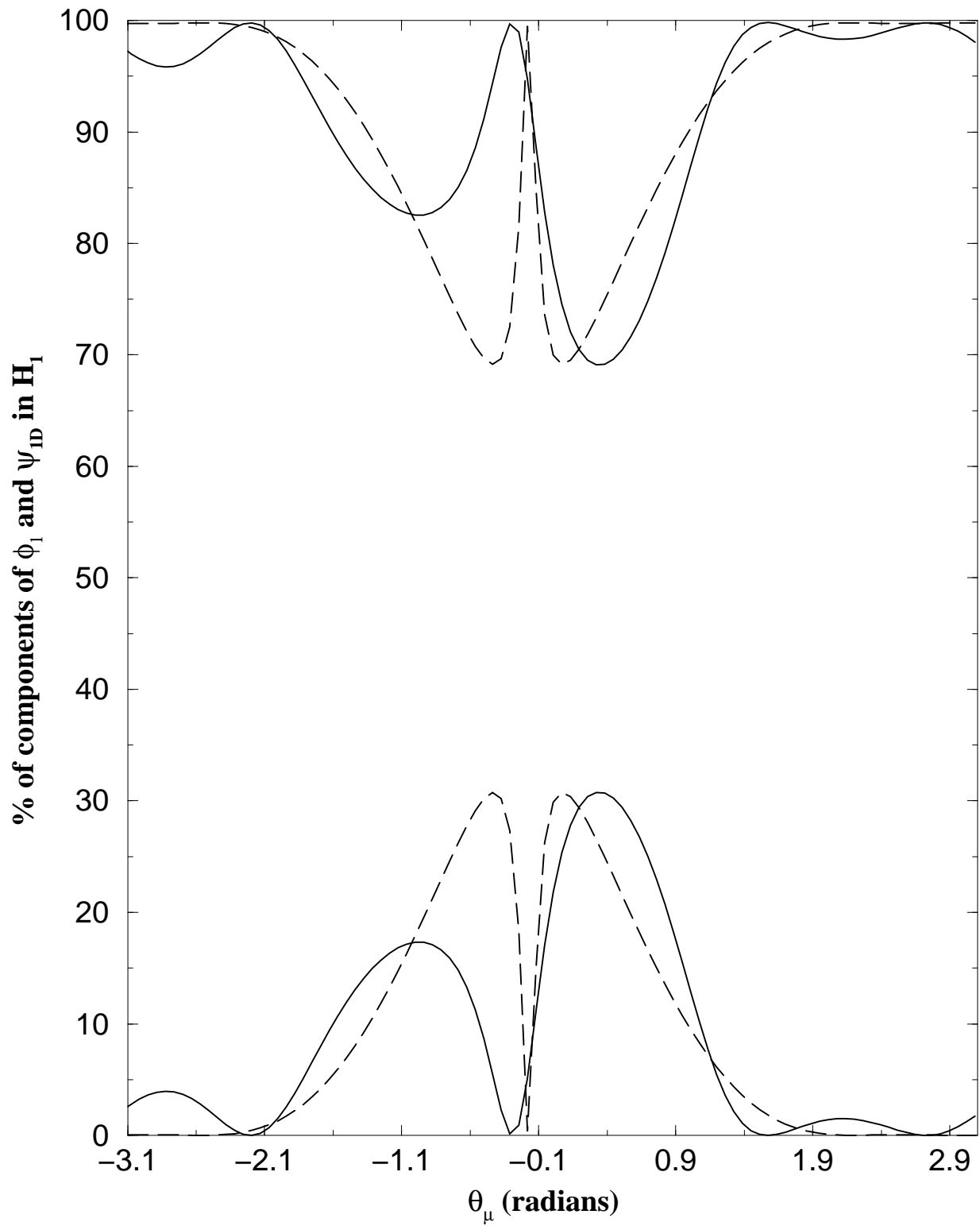
\end{document}